# Cross-sectional imaging of individual layers and buried interfaces of graphene-based heterostructures and superlattices

S. J. Haigh[1]*, A. Gholinia[1], R. Jalil[2], S. Romani[3], L. Britnell[2], D.C. Elias[2], K. S. Novoselov[2], L. A. Ponomarenko[2], A. K. Geim[2], R. Gorbachev[2]*

[1] School of Materials and [2] Manchester Centre for Mesoscience & Nanotechnology,
University of Manchester, Manchester M13 9PL, UK
[3] Department of Engineering, University of Liverpool, Liverpool L69 3GH, UK

**By stacking various two-dimensional (2D) atomic crystals [1] on top of each other, it is possible to create multilayer heterostructures and devices with designed electronic properties [2-5]. However, various adsorbates become trapped between layers during their assembly, and this not only affects the resulting quality but also prevents the formation of a true artificial layered crystal upheld by van der Waals interaction, creating instead a laminate glued together by contamination. Transmission electron microscopy (TEM) has shown that graphene and boron nitride monolayers, the two best characterized 2D crystals, are densely covered with hydrocarbons (even after thermal annealing in high vacuum) and exhibit only small clean patches suitable for atomic resolution imaging [6-10]. This observation seems detrimental for any realistic prospect of creating van der Waals materials and heterostructures with atomically sharp interfaces. Here we employ cross sectional TEM to take a side view of several graphene–boron nitride heterostructures. We find that the trapped hydrocarbons segregate into isolated pockets, leaving the interfaces atomically clean. Moreover, we observe a clear correlation between interface roughness and the electronic quality of encapsulated graphene. This work proves the concept of heterostructures assembled with atomic layer precision and provides their first TEM images.**

Top view TEM of graphene and boron nitride monolayers has allowed visualization and analysis of various types of defects including ripples, vacancies, substitutional atoms, adatoms, grain boundaries and edges [6-14]. Elemental analysis at the level of individual atoms can now be achieved by means of annular dark field imaging [15], electron energy-loss spectroscopy [7,9,10,16,17] or energy dispersive x-ray spectroscopy [18]. These atomic-scale insights are important for fundamental and technological progress in this research area because defects determine the electronic quality of graphene devices [19] and can lead to pinholes in tunnel barriers [20-22].

Most recently, the field has expanded beyond studying graphene and isolated 2D crystals. There is rapidly growing interest in atomic-scale heterostructures made from a combination of alternating layers of graphene, hexagonal boron-nitride (hBN), $MoS_2$, etc. Such heterostructures provide a higher electronic quality for lateral graphene devices [23,24] and, also, allow a conceptually new degree of flexibility in designing electronic, optoelectronic, micromechanical and other devices [2-5]. New fundamental physics not present for individual 2D crystals is widely expected to emerge in heterostructures with atomically thin barriers and quantum wells [3,25]. However, defects and, in particular, adsorbates trapped between 2D crystals can diminish their quality. In the case of multilayer structures, their top view TEM images are difficult or impossible to interpret because many different layers are superimposed in projection. Conventional surface-analysis techniques are also insensitive to the buried interfaces and trapped contamination. To obtain information about such atomic-scale heterostructures, a different approach is necessary.

In this report, we have studied multilayer heterostructure devices in which mono- and bi- layer graphene crystals were individually contacted and interlaid between atomically thin hBN crystals. After microfabrication (see Methods) and electron transport measurements, TEM specimens were extracted from the studied devices as follows. A suitable region for side view imaging was chosen by using optical and scanning electron microscopy (SEM) (Fig 1). An Au-Pd film (50 nm thick) was then sputtered onto the whole surface, followed by a narrow Pt strap (1 μm thick) deposited over the region of interest. The metal layers served to prevent charging and protect the selected region from ion damage. A focused ion beam (FIB) milling was then used to produce a cross sectional slice under the Pt strap. The slice was then transferred and attached to a TEM grid, a sample preparation approach known as the lift-out method [26,27]. Finally, we employed low energy ion milling to thin the extracted slices down to 20-70 nm so that they could be analyzed at high resolution by scanning transmission electron microscopy (STEM) [28,29].

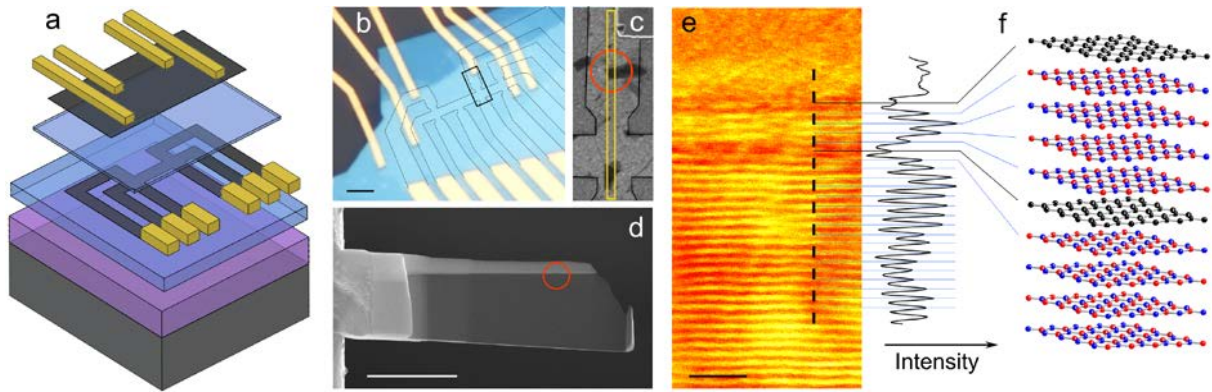

**Figure 1. Cross-sectional TEM of graphene-hBN heterostructures**. **a** – Schematic of one of our devices: two graphene monolayers (dark gray) are interlaid with hBN crystals (blue) and have separate electrical contacts (yellow). The heterostructure is fabricated on top of an oxidized silicon wafer (violet and gray). **b** – Optical image of the device before extracting a cross sectional sample. The graphene Hall bar is highlighted with thin black lines. The black rectangle (1.8 µm x 4.4 µm) indicates the area enlarged in the SEM micrograph (**c**), in which the yellow rectangle marks the region chosen for our study. **d** – Low magnification SEM image of the final cross sectional slice for TEM imaging. Red circles in **c, d** indicate a 'bubble' of trapped adsorbates. **e** – High resolution bright field aberration corrected STEM image of the graphene-hBN heterostructure in the circled region. **f** – Averaged line scan of the intensity along the dashed black line in **e** (total length 9.4 nm, integration width 2 nm) with a schematic of the atomic layer sequence for this particular device shown on the right in which carbon, boron and nitrogen atoms are represented as black, blue and red respectively. Scale bars in (b) and (d) are 3µm. Scale bar in (e) is 2nm.

Fig. 1 shows one of our investigated devices and its step by step preparation for STEM. The particular device consists of monolayer graphene placed on top of a relatively thick (30 nm) hBN crystal, covered with a few-layer hBN spacer and then another graphene sheet on top (Fig. 1a). A cross sectional specimen extracted from the device is shown in Fig. 1d. At this low magnification only the Pt protective layer and Si substrate are visible. Individual atomic layers become clearly resolved by using high resolution STEM. One can easily count them in Fig. 1e. In this projection, little difference is expected between graphene and hBN atomic planes but our images consistently show two regions of slightly reduced contrast and intensity, which coincide with the expected positions of the two graphene sheets, above and below the hBN spacer. We attribute this to higher susceptibility of graphene or the interfaces to ion beam damage. The dips in contrast allow us to estimate the hBN spacer thickness, $d$, as 4 atomic layers, in agreement with atomic force microscopy (AFM) measurements during fabrication of the device ($d \approx$1-2 nm) and its tunneling characteristics yielding $d$ =4 hBN layers [22].

Because the individual crystals are prepared under ambient conditions, various chemical species get absorbed on their surfaces, as commonly seen by SEM and top-view TEM [6-10]. Note that AFM and scanning tunneling microscopy usually do not 'see' this contamination, probably because it is easily moved by the tip. During fabrication of our heterostructures, the contamination layer is expected to remain between the layers. In this work, we find that adsorbates tend to diffuse over micron scale distances and form relatively large (submicron) pockets of the trapped material, which are clearly seen as 'bubbles' in an optical microscope and as dark spots in SEM (Fig. 1c and Supplementary Information). These inhomogeneities can be detrimental to device performance and decrease graphene's carrier mobility, $\mu$. To avoid this, we normally heated our devices to 300°C to enhance the diffusion. This helped to obtain clean, atomically flat areas large enough to fabricate graphene devices of a micron width with no bubbles within the active area (see Supplementary Information for our studies on bubbles' morphology). To find out what is inside the bubbles, so as to identify their chemistry and origins, we have deliberately chosen in Fig. 1 to image a cross-section that goes through one of the bubbles. Another example is shown in Fig. 2 where a low magnification image provides a better view of the contamination bubble. The bubble contains an amorphous material that is trapped between the top graphene layer ($G_T$) and hBN spacer ($BN_T$) and becomes hardened, probably during ion milling (Supplementary Information). Chemical analysis using energy dispersive x-ray (EDX) and electron energy loss (EEL) spectroscopy has revealed that the trapped contaminants are mostly hydrocarbons (Fig. 2b).

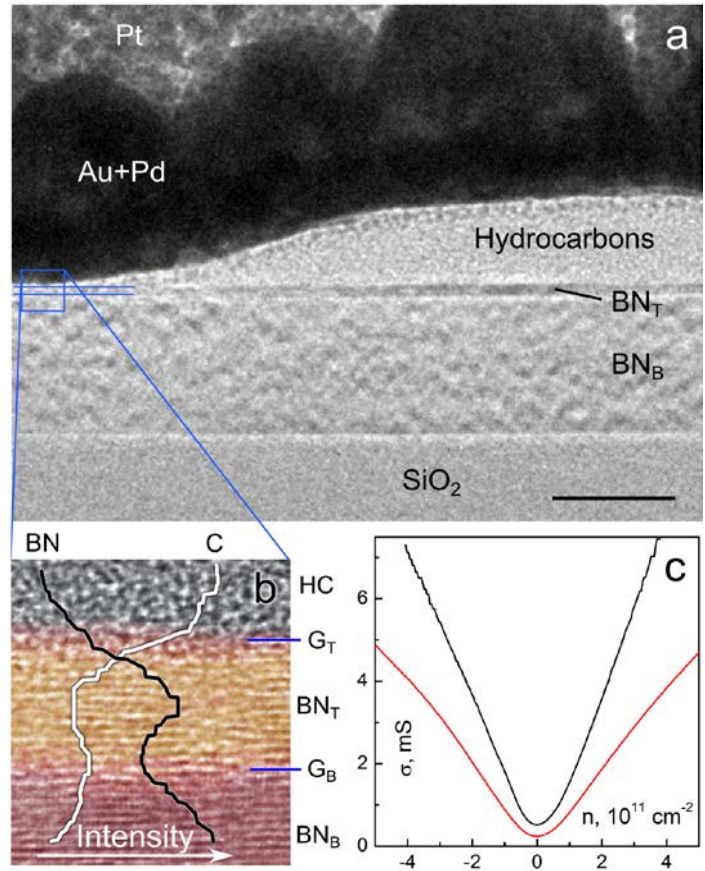

**Figure 2. Characterization of chemical species trapped between layers**. **a** – bright field TEM image showing the top part of the oxidized Si wafer, BN substrate layer ($BN_B$), BN spacer ($BN_T$), hydrocarbon residue (HC) and metallic coating (Au-Pd layer + Pt strap). Scale bar is 25 nm. **b** – Bright field aberration corrected STEM image of an enlarged region of the same heterostructure. Superimposed are elemental profiles for the carbon-K edge and boron-K edge extracted from EEL spectra acquired sequentially in the STEM mode. The peak in the C content for the bottom graphene layer (GB) corresponds to a maximum relative atomic composition of 55%. The image is 10 nm wide. **c** – Electrical characterization of graphene layers in this device. Conductivity σ of top (red) and bottom (black) graphene as a function of carrier density *n*. The larger the slope of σ(*n*), the higher graphene's μ.

The acquired cross sectional EEL spectra also allow us to chemically identify approximate locations of the graphene layers. Indeed, Fig. 2b shows a drop in the BN signal and a simultaneous increase in the carbon signal (note that on this particular image $G_T$ is absent, being lifted away by hydrocarbon contamination). The limited resolution (≈1 nm) of our elemental mapping in Fig. 2b is consistent with the spreading of the focused electron probe, which can be estimated from the thickness of the cross sectional slice (≈40 nm) and the employed 20 mrad convergence angle [30]. For the device in Fig. 2 the EEL spectroscopy yields *d* =4-5 nm for $BN_T$, in agreement with the device's AFM characterization and its quantum capacitance measurements [3]. For such relatively thick hBN spacers, the $BN_T$ and $BN_B$ layers are often observed to have different contrast in the medium angle annular dark field STEM images (shown in false color in Fig. 2b; $d \approx$4.5 nm), which is attributed to these hBN crystals having different lattice orientations, with the $BN_T$ rotated about the c-axis relative to $BN_B$.

The isolated pockets of trapped hydrocarbons are of concern for fabrication and possible lateral sizes of graphene-hBN heterostructures but even more serious is the question of whether such contamination remains at the interfaces in a sub-monolayer thickness, which can affect their electronic quality and smoothness. We address this issue by analyzing our high-resolution TEM images. To this end, we have used an analysis of peak positions in the intensity spectra (such as in Fig. 1f) to accurately measure interlayer distances (see Supplementary Information). We find that within experimental accuracy the spacing between hBN and graphene planes is indistinguishable from the basal plane separation in bulk hBN (for the device in Fig. 1 our accuracy given by a standard deviation $\Delta$ was $\approx$0.07 nm). This suggests that the interface between graphene and hBN cannot contain more than a small fraction of a monolayer of adsorbates, that is, the interface is practically clean and atomically sharp.

Furthermore, we have estimated the roughness of each atomic layer in our TEM images by analyzing the standard deviation of intensity peak positions from a straight line (Supplementary Information). This analysis reveals a significant difference in the average roughness $\delta$ between the top and bottom graphene planes ($\delta \approx$0.072 and 0.038 nm for $G_T$ and $G_B$, respectively, for the device in Fig. 1). For comparison, atomic planes within $BN_T$ and $BN_B$ yield the same $\delta$ as $G_B$ with the same experimental error of ±0.023 nm. We believe that, despite the use of low energy ion polishing and short dwell times during STEM imaging, the background level of roughness can be due to beam damage. This does not account for the 3 times higher $\delta$ for the top graphene layer. Moreover, the observed roughness correlates with graphene's electronic quality so that $G_T$ and $G_B$ exhibited $\mu \approx$60,000 and 120,000 cm$^2$/Vs, respectively, for the device in Fig. 2c. Lower $\mu$ in $G_T$ was a systematic effect observed for all our devices. We speculate that the difference in $\mu$ is a result of stronger rippling [31] for the unprotected top graphene or of more adsorbates trapped at the top interface. In either case, this observation shows that the encapsulation is important for achieving atomically flat and high-$\mu$ graphene, in agreement with results of ref. [24].

Presented in Fig. 3 is an example of a more sophisticated heterostructure, comprising several graphene bilayers separated by hBN bilayers. Fabrication of this device required sequential deposition of many crystals on top of each other (we made devices with up to 10 different layers). AFM and optical characterization of each bilayer before the assembly have assured the exact heterostructure composition as designed. For the device in Fig. 3, its high resolution bright field and high angle annular dark field (HAADF) imaging reveal a region of different contrast for the ten atomic layers representing the superlattice region. Compositional analysis using EEL spectroscopy shows a systematic increase in the carbon K-edge signal within this region but is unable to identify the individual graphene bilayers because the resolution is limited to ~1 nm. Analysis of the intensity profiles extracted from HAADF STEM images (like that shown in Fig 3b) reveals a spacing for the hBN-graphene superlattice of ≈0.332 ±0.043nm, equal to the interlayer distance in bulk hBN (measured to be ≈0.326nm with the same accuracy for the upper part of the BN substrate). No systematic change in lattice separation was observed as a function of distance from the hBN substrate (Supplementary Information). These measurements prove again that the graphene-hBN interfaces become atomically clean due to the segregation of adsorbed hydrocarbons.

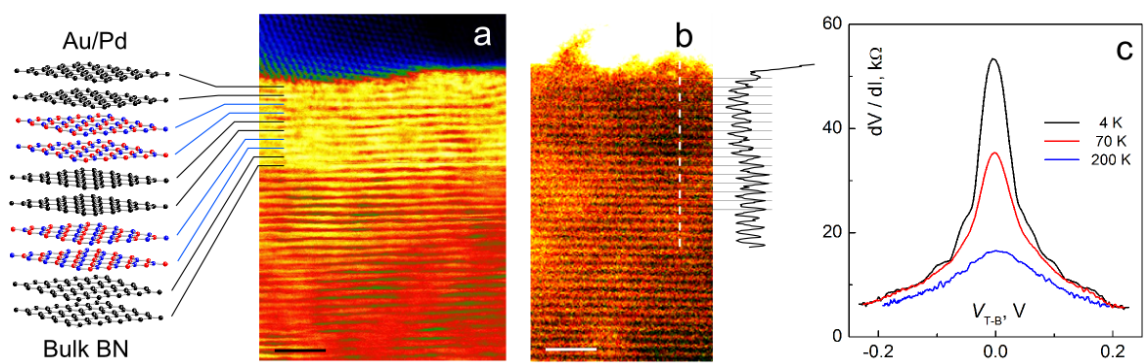

**Figure 3. Graphene-BN superlattice. a** – Bright field cross sectional STEM of a stack of graphene and hBN bilayers with the layer sequence schematically shown to the left. **b** – HAADF STEM image of the same superlattice with an intensity line profile of length 7 nm and integration width 5 nm plotted to the right. **c** – Vertical transport through the superlattice. Current, $I$, was measured as a function of bias voltage, $V_{T-B}$, applied between the top and bottom graphene bilayers. The differential resistance $dV/dI$ exhibits a peak at zero bias which grows with decreasing temperature. Scale bars are 2 nm.

The structural TEM studies also aid interpretation of electron transport measurements. For the described graphene-BN superlattice, metallic contacts were attached to the top and bottom graphene bilayers to probe electrical conductivity in the vertical direction. Fig. 3c shows differential resistivity for the same superlattice as imaged in Figs. 3a,b. The measured I-V characteristics are strongly nonlinear because hBN bilayers act as tunnel barriers, in agreement with the recent studies of tunneling through individual hBN crystals [20-22]. On the other hand, the nonlinearity appears in the superlattice at biases as small as <10 meV and the zero-bias peak grows by a factor of 5 between room and liquid-helium temperatures. This behavior is inconsistent with that observed for individual hBN bilayers [22]. Its full interpretation is beyond the scope of the present report. Let us only mention that we attribute the zero-bias peak to a combination of two effects: a decrease in the tunneling density of states at the neutrality point in bilayer graphene as temperature decreases and a 'zero-bias anomaly' that can appear due to the interaction of a tunneling electron with a virtual hole left behind in a 2D electronic system [32].

In conclusion, our high-resolution images show that new kinds of heterostructures with atomically sharp interfaces can be assembled in various combinations from the available library of 2D crystals. It is highly nontrivial that the contamination inevitably present on top of 2D crystals and trapped during their assembly segregates into isolated pockets, leaving the buried interfaces clean and atomically flat. In addition, our work shows that cross sectional TEM can be a useful tool to aid interpretation of transport measurements and therefore further progress in graphene-based electronics, allowing side view characterization with atomic resolution.

**METHODS**

**Fabrication of multilayer heterostructure devices**

The stack begins with a thick (~50 nm) hBN flake deposited on top of an oxidized Si wafer. Each subsequent layer is prepared on a separate wafer, delaminated from the surface and transferred on top of the target crystal. The freshly deposited layer can then be shaped by reactive plasma etching and annealed (300° C in Ar/H$_2$) in order to remove processing residues. Because the graphene layers are interlaid by insulating hBN spacers, each of them can be separately connected (we used Au/Ti contacts) and studied independently. More details on device fabrication can be found in ref. [3]. For additional information on contamination pockets (bubbles), see Supplementary Information.

**Preparation of TEM samples**

A dual beam instrument (FEI Nova NanoLab 600) has been used for site specific preparation of cross sectional samples suitable for TEM analysis using the lift-out approach [26-29]. This instrument combines a focused ion beam (FIB) and a scanning electron microscope (SEM) column into the same chamber and is also fitted with a gas-injection system to allow local material deposition and material-specific preferential milling to be performed by introducing reactive gases in the vicinity of the electron or ion probe. The electron column delivers the imaging abilities of the SEM and is at the same time less destructive than FIB imaging. SEM imaging of the device prior to milling allows one to identify an area suitable for side view imaging. After sputtering of a 50 nm Au- Pd coating on the whole surface ex-situ, the Au/Ti contacts on graphene were still visible as raised regions in the secondary electron image. These were used to correctly position the ion beam so that a Pt strap layer could be deposited on the surface at a chosen location, increasing the metallic layer above the device to ~1 μm. The strap protects the region of interest during milling as well as providing mechanical stability to the cross sectional slice after its removal. Trenches were milled around the strap by using a 30 kV Ga$^+$ beam with a current of 0.1- 5nA. Before removing the final edge supporting the milled slice and milling beneath it to free from the substrate, one end of the Pt strap slice was welded to a nanomanipulator needle using further Pt deposition. The cross sectional slice with typical dimensions of 1 μm x 5 μm x 10 μm could then be extracted and transferred to an Omniprobe copper half grid as required for TEM. The slice was then welded onto the grid using further Pt deposition so that it could be safely separated from the nanomanipulator by FIB milling. A final gentle polish with Ga+ ions (at 5kV and 50pA) was used to remove side damage and reduce the specimen thickness to 20-70nm as measured using energy filtered TEM. The fact that the cross sectional slice was precisely extracted from the chosen spot was confirmed for all devices by comparing the positions of identifiable features such as Au contacts and /or hydrocarbon bubbles, which are visible both in the SEM images of the original device and within TEM images of the prepared cross section (see, e.g., Fig. 1).

**Transmission electron microscopy and image analysis**

TEM imaging was carried out using a Tecnai F30 FEG-TEM operated at 300 kV and equipped with an energy dispersive X-ray detector. STEM imaging was performed by using a probe-side aberration-corrected JEOL 2100F operated at 200 kV with the third order spherical aberration set to zero (±5 µm). The multilayer structures were oriented along an <hkl0> crystallographic direction by taking advantage of the Kukuchi bands of the Si substrate. Images were collected using a convergence angle of 20 mrad and a high angle annular dark field (HAADF) detector with an inner (outer) collection angle of 74 (196) mrad. Electron energy loss (EEL) spectroscopy data was acquired using a Quantum Gatan Imaging Filter and a collection angle of 24 mrad. Analysis of the EELS line scan data was performed using Digital Micrograph software with pre and post edge windows of ~20 eV.

Analysis of the local interlayer separation and layer roughness was performed using line profiles extracted from the images within the Semper image processing software. Details of the analysis are provided in the Supplementary Information.

# SUPPLEMENTARY INFORMATION

**Contamination bubbles trapped at the graphene-hBN interface**

The bubbles described in the main text are commonly seen after the transfer of graphene or monolayer hBN on top of an atomically flat substrate, regardless whether it is graphite or hBN. The bubbles occur even when a solvent free technique is used [S1], in which case neither of the two attaching surfaces has been in contact with a liquid. Figure S1a shows that the bubbles occur only within the attachment area and cannot be seen anywhere else on the surface, which suggests that a material is trapped between the attached crystals. Depending on the conditions of the transfer procedure, the distance between bubbles may vary between, typically, 1 and 20 µm (Fig. S1). After annealing at 200 C, the bubbles tend to aggregate into bigger ones spaced further apart so that it is easier to find sufficiently large bubble-free regions for device fabrication. Fig. S1b shows a scanning electron micrograph (SEM) of several bubbles trapped under graphene on top an hBN substrate. They are connected by narrow creases that apparently appear because of the strain associated with the enclosures. Scanning Raman microscopy shows that graphene is notably stretched at bubbles' locations to surround some trapped material [S2].

Initial information about what is inside such bubbles has been obtained by using atomic force microscopy (AFM). Figure S1c shows an AFM topography image of a bubble trapped at the interface of a graphite substrate and a few-layer hBN crystal. The color scale is chosen to show interference-like contours, and the bubble is approximately 50 nm high. The substrate extends beyond the image boundaries whereas terraces in the hBN are visible as three planar regions that correspond to mono-, bi- and tri- layer hBN. The bubble is trapped at the mono-/bi- layer boundary. This image is chosen because it shows that due to lower stiffness of the hBN monolayer as compared to the bilayer, the bubble's content is squeezed preferentially into the monolayer region, which results in the asymmetric shape. It was impossible to move or reshape such bubbles with an AFM tip. Also, upon indentation, the bubbles relaxed to their original shape after tip retraction, unless the force was sufficient to break the top layer. Figure S1d shows the same bubble after the AFM probe was used to scratch a line in the monolayer hBN region. The resultant damage is clearly visible: A part of the bubble's content escaped and the bubble re-sealed itself. One can also see that creases appeared due to changes in the shape of the surviving bubble. There is no sign of the escaped material on the surrounding surface, suggesting that the contamination is either volatile or, more likely, highly mobile under the AFM tip.

It is worth mentioning that bubbles can sometimes appear at the interface between graphene and the $SiO_2$ surface if graphene crystals are large, >100 micrometers in size [S2,S3]. However, those bubbles contain a gas as clearly seen through their compressibility by an AFM tip or by applying an electric field [S3]. The different content is presumably related to a highly porous surface of $SiO_2$ as compared to atomically flat hBN or graphite. The difference between the two kinds of bubbles is most clearly seen after plasma etching or FIB irradiation. In the case of $SiO_2$, after a top graphene layer is etched away, the remaining surface becomes pristine with nothing left at bubbles' positions. In contrast, similar exposure of bubbles at graphene/hBN interfaces always leaves some immobile hard material on the surface in the same positions where the bubbles were. AFM and SEM analysis indicates that the residue is mostly likely to be cross-linked hydrocarbons, in agreement with the TEM analysis presented in the main text.

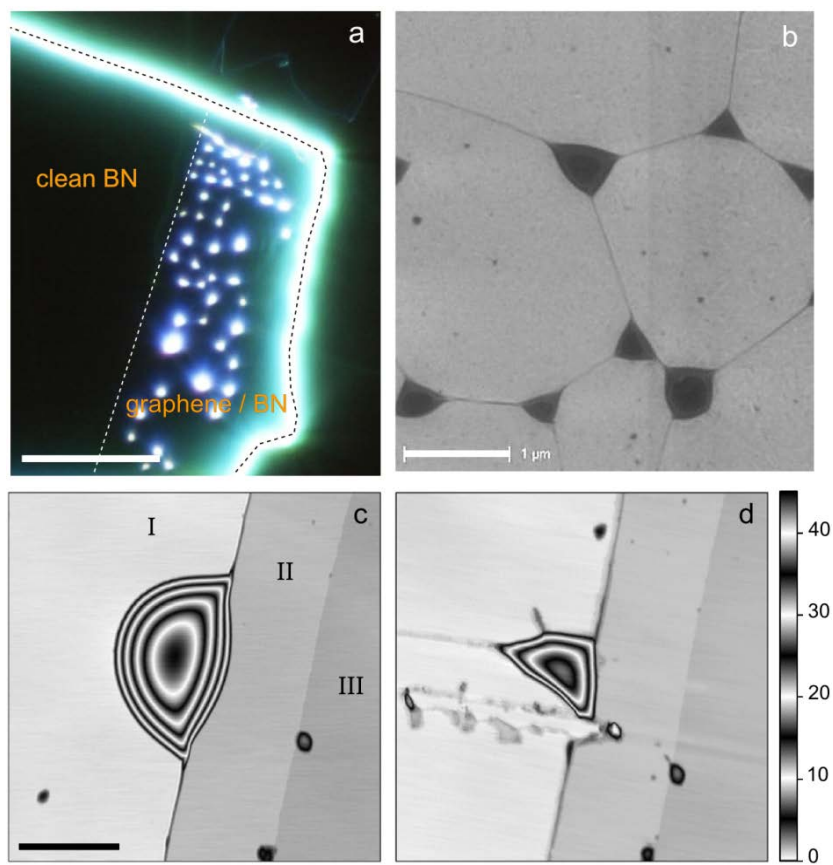

**Figure S1.** Hydrocarbon bubbles at the graphene-hBN interface. **(a)** Dark field optical micrograph of a BN crystal (~50 nm thick) outlined with the black dashed curve, which is partially covered with a graphene monolayer as outlined with the white line. The contamination bubbles are clearly visible as bright spots. The scale bar is 20 μm. **(b)** SEM image of monolayer graphene on top of hBN. Bubbles give the pronounced dark contrast and are interconnected with creases. Scale bar - 1 μm. **(c)** AFM image of a bubble trapped between a graphite surface and an hBN flake with single (I), double (II) and triple (III) layer regions from left to right, respectively. **(d)** The same bubble after tearing hBN with an AFM tip. The scratch is clearly visible. The scale bar for (c) and (d) is 500 nm.

**Image analysis: Interlayer separation**

For analysis of the interlayer separation, line profile intensity spectra were obtained from raw images using SEMPER image processing software [S4]. The individual spectra were extracted perpendicular to the graphene-hBN interface using bilinear interpolation between the four neighboring source pixels. The peak positions corresponding to the individual atomic planes within each spectrum were first identified and then refined using a local center of mass algorithm [S4]. The interlayer separation was extracted as the difference in peak position for each consecutive peak pair.

Figure S2a shows examples of such analysis for the superlattice device presented in Fig 3 of the main text. Dot colors correspond to 4 different spectra (length 16 nm, pixel size 0.004 nm, integration width 4 nm). For a reference, Fig S2a also shows an individual HAADF intensity spectrum with the dashed lines indicating the average periodicity. The scatter of the data points is notably larger in the superlattice region (mostly due to the weaker contrast of the peaks; see Fig S2a) but no systematic change in the lattice spacing is noticeable between bulk hBN on the right and the superlattice region on the left. This analysis was carried out for different areas in the sample, and the results were same within one standard deviation.

The above analysis has also been done for the double-layer graphene device presented in Fig 1 of the main text. In this case, 20 line spectra (length 10 nm, pixel size 0.002 nm, integration width 0.4 nm) were obtained for different regions of the sample. The interlayer separations next to the encapsulated graphene layer were compared to those in the center of the hBN spacer and to those within the hBN substrate. Within scatter, no difference in the interlayer separation was noticed. Because the encapsulated graphene layer has weaker contrast compared to that of the surrounding hBN, we have also analyzed the distance between the two nearest hBN layers above and below of the bottom graphene layer ($G_B$) and observed double of the expected bulk BN value. This strengthens further the conclusion that the interlayer separation within graphene-hBN heterostructures is little different from that in bulk hBN or graphite.

**Image analysis: Roughness**

For the roughness analysis we used intensity spectra (length 10 nm, pixel size 0.002 nm, integration width 0.4 nm) which were obtained sequentially at regular 0.4nm intervals along the interface (for an example of individual spectra, see Fig S2b, right). For each set of spectra we traced the position of a selected layer on each spectrum and calculated the mean peak position value. Then, the roughness of each atomic layer could be evaluated by comparing the local peak positions in each spectrum to the mean position. The bar graph in Figure S2b plots the found deviations in the individual peak positions from the mean value for the top (grey) and bottom (orange) graphene layers extracted from 20 individual spectra such as shown on the right. This figure illustrates that the top graphene layer is rougher than the bottom one. In addition, both layers seem to follow some overall landscape. Note that the roughness in the beam direction cannot be determined in our experiment and, therefore, the found roughness values can be used only as an indication of the sample morphology.

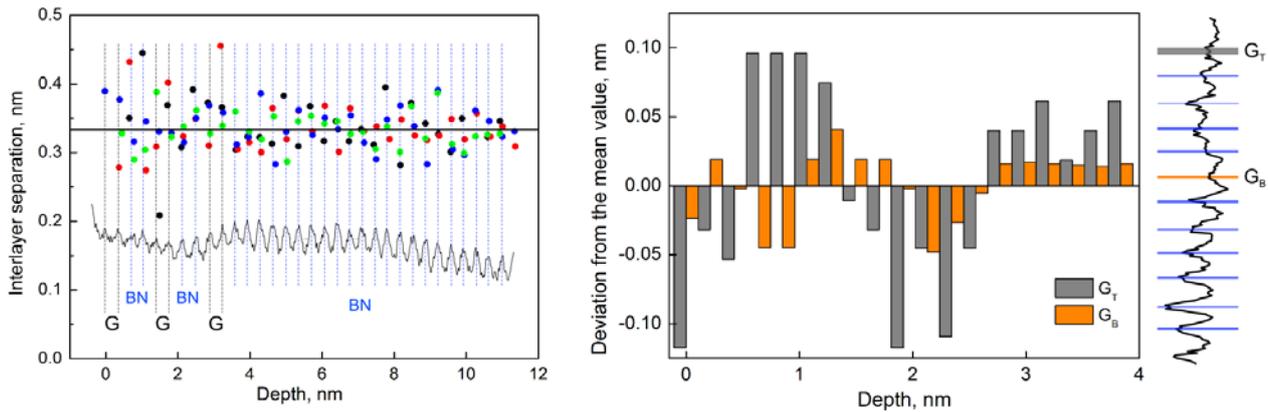

**Figure S2.** Image analysis. (**a**) Interlayer separation was evaluated as a difference in peak positions within individual line spectra such as shown in the lower part of the figure. Each dot plots the distance between two nearest peaks in such spectra. Different colors represent 4 different line scans. Many more scans were analyzed. The horizontal line corresponds to the interlayer separation in bulk hBN (0.33nm). The data are from the superlattice device in Fig 3. (**b**) Bar graph showing deviation of the individual peak positions from the mean value for the top $G_T$ and bottom $G_B$ graphene layers, indicated by color in a representative line spectra on the right. The data corresponds to the double-layer graphene device in Fig. 1.

**Supplementary References**

[S1] Ponomarenko, L. A. *et al.* Tunable metal-insulator transition in double-layer graphene heterostructures. *Nature Phys.* **7,** 958-961 (2011).

[S2]. Zabel, J. *et al.* Raman Spectroscopy of Graphene and Bilayer under Biaxial Strain: Bubbles and Balloons. *Nano Lett.* **12**, 617 (2012).

[S3] Georgiou, T. *et al.* Graphene bubbles with controllable curvature. *App. Phys. Lett.* **99**, 093103 (2011).

[S4] Saxton, W.O., T.J. Pitt, and M. Horner, Digital image processing: The semper system. *Ultramicroscopy* **4,** 343 (1979).